\documentstyle[12pt]{article}
\input epsf
\begin{document}
\title{
\hfill{\small TPBU-10-95}
\\
\hfill{\small cond-mat/9512155}
\\
\hfill{\small December 1995}
\\
\vspace*{0.5cm} \sc Aspect ratio analysis for ground states \\
of bosons in  anisotropic traps
\vspace*{0.3cm}}
\author{{\sc Kirill N. Ilinski}
${}^{1,2}$\ and {\sc Alexander Moroz}${}^{1,3}$}
\date{
\protect\normalsize
\it ${}^{1}$School of Physics and Space Research, The University of
Birmingham, Birmingham B15 2TT, United Kingdom\\
${}^{2}$Institute of Spectroscopy, Russian Academy of Sciences, Troitsk, Moscow region, 142092, Russian Federation\\
${}^{3}$Institute of Physics, Na Slovance 2, CZ-180 40 Praha 8, 
Czech Republic}
%
%\vspace*{0.3cm}
%
\maketitle
\begin{center}
{\large\sc abstract}
\end{center}
%\begin{abstract}
Characteristics of the initial condensate in 
the recent experiment on Bose-Einstein condensation (BEC)
of ${}^{87}$Rb atoms in an anisotropic magnetic trap is discussed.
Given the aspect ratio $R$, the quality of BEC is estimated.
A simple analytical Ansatz for the initial condensate wave function
is proposed as a function of the aspect ratio
which, in contrast to the Baym-Pethick trial wave function,
can be used for any interaction strength,
reproduces both the weak and the strong interaction limits, 
and which is in better agreement with numerical results than the latter.
%\end{abstract}

\vspace*{0.2cm}

\vspace{0.6cm}
%\pacs{PACS numbers : 03.75.Fi, 05.30.Jp, 32.80.Pj }
%\narrowtext

\vspace{1.2cm}

\begin{center}
{\bf (to appear in J. Res. Nat. Inst. of Standards and Technology)}
\end{center}

\thispagestyle{empty}
\baselineskip 20pt
\newpage
\setcounter{page}{1}
\section{Introduction}
%%%%%%%%%%%%%%%
Bose-Einstein condensation (BEC)  is a phenomenon
where a macroscopic number of particles is in the ground state
of the system at finite temperature. The phenomenon  of BEC plays
significant roles in many branches of physics \cite{GSS}.
Because of the presence of strong interactions,
BEC has been  inferred rather than directly observed so far.
Recently, however, three different groups have reported the 
direct evidence of BEC in weakly interacting systems of
atoms  such as rubidium \cite{Exp1}, lithium \cite{Exp2},
and sodium \cite{Exp3}, confined in
anisotropic magnetic traps and cooled down to very low temperatures.
These experiments show promise of becoming a new
laboratory for quantum statistical phenomena
that are inaccessible to other conventional techniques and of enabling
us to experimental study  phenomena
that have been addressed only theoretically, such as spontaneous symmetry
breaking and decay of unstable macroscopic states.
They may also advance our understanding of superconductivity and
superfluidity in more complex systems. Moreover, the technology
used in the experiments has the possibility to be extended to
produce a veritable atomic laser that is bound to  have many
applications in pure science and technology \cite{Bur}.

In recent experiments in the system of  rubidium atoms \cite{Exp1} and  
sodium atoms \cite{Exp3},
the onset of BEC is signalled by a narrow peak on top of a broad
thermal velocity distribution centered at zero velocity.
This peak exhibits the nonthermal, anisotropic velocity distribution expected
of the minimum-energy quantum state of the magnetic trap in contrast
to the  thermal, isotropic velocity distribution observed in the
broad uncondensed fraction. The parameter which characterizes the
asymmetry of the velocity distribution function is the so called {\em aspect
ratio} $R\equiv \sqrt{\langle p_z^2\rangle/\langle p_x^2\rangle}$.
In the experiment of Anderson et al.  \cite{Exp1},
rubidium atoms are initially trapped and cooled down in a strong 
magnetic trap which can be described as a
three-dimensional ($3D$) harmonic potential
cylindrically symmetric about the $z$-axis, with tunable frequency
$\omega _{z}$ (in the $z$ direction) and
$\omega _{\perp}=\omega _{z}/\lambda$ (in the $xy$-plane), with the
asymmetry parameter $\lambda=\sqrt{8}$.
The corresponding oscillator lengths are
$
a_{\perp (z)} = (\hbar/m \omega _{\perp (z)})^{1/2}
=  1.25  (0.74) \times 10^{-4} \mbox{cm},
$
where $m$ is the atomic mass. After some time, the cloud of rubidium atoms 
is adiabatically released to  a weaker magnetic trap whose spring constants 
are 10 times weaker than when BEC forms.
The condensate is then examined after ballistic expansion from the weak trap.
The ballistic expansion is properly modelled numerically by a self-consistent
wavefunction  calculated by  Holland and Cooper
\cite{HC} which generalizes  previous investigation of 
the symmetric evolution \cite{RHB}.
 
Similarly to previous recent studies \cite{BP,EDC,DS}, 
we shall confine ourselves to the system of
${}^{87}$Rb atoms and we will discuss the characteristics of 
the initial condensate formed. Although the characteristics  of 
the initial condensate have not been measured directly yet, 
there is hope that it will be possible so in future experiments.

A theoretical picture of the initial 
condensate was produced 
within the  Hartree-Fock (HF) approximation \cite{BP,EDC,DS} using the
Ginzburg-Pitaevskii-Gross (GPG)  energy functional \cite{GP} and
associated with it the Nonlinear Schr\"{o}dinger Equation (NSE).
Excitations, using the technique of the Bogoliubov transformation,
have been described by Fetter~\cite{F}.
Baym and Pethick \cite{BP} have gained an insight into the problem
by assuming that, similarly to  the noninteracting case,
a gaussian gives a reasonable variational ground-state wave function, 
the only effect of interactions being a renormalization of
oscillator frequencies. They show  that the first
effect of interactions is to reduce the density of the cloud of
particles in the central region from the free particle situation and
expand it in the transverse direction. These qualitative
features were confirmed by  Edwards et al \cite{EDC} and Dalfovo  and Stringari~\cite{DS} by solving the NSE numerically. 

\begin{figure}
\centerline{\epsfxsize=10cm\epsfbox{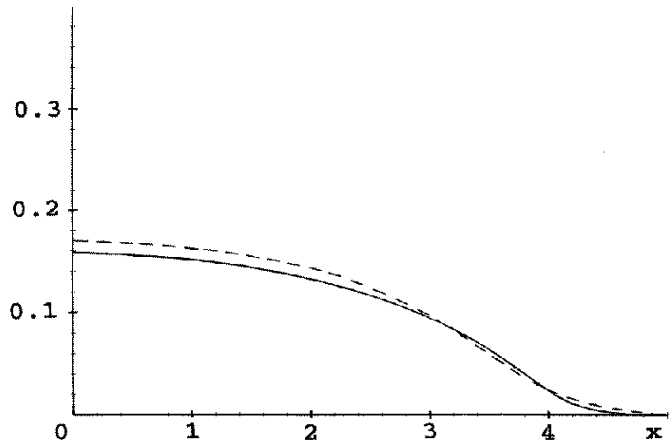}}
\caption{A comparison of the $x$-dependence of the numerical solution 
of NSE for the ground state [10] (dashed line) 
and our approximate solution (solid line) in the case of 
$N=5000$ atoms of ${}^{87}$Rb, when $A=502$ and $C_{ansatz}=2.2$. 
The value of the aspect ratio, $R(A,\lambda)=2.3$, is taken 
from Ref\ [10]. } 
\label{pl3}
\end{figure}

In the present paper it is shown that  although, qualitatively, 
the Baym and Pethick (BP) scenario \cite{BP} is correct, nevertheless, 
the BP wave function does
not describe BEC regime well. As we show in the next section, this can 
be explained by the fact that the BP
wave function is actually the square root of the first order
density function in the high temperature expansion of the partition
function of the $\delta$-function interacting Bose gas in
the kinetic energy, and hence describes  high temperature
properties of the system. Higher order corrections are needed
to obtain low temperature properties such as BEC in agreement
with the numerics. In particular, this explains the unreasonably high 
aspect ratio in BP estimations (up to $4.2$, i.e., $2.5$ higher than in
the noninteracting case). All these prompt a search for another 
trial variational function for the ground state. In the section 
\ref{ans} we derive a simple analytical Ansatz 
[see Eq. (\ref{fe}) and Fig. \ref{pl3}] which, for a given numerical
value of the aspect ratio, describes well the ground state
properties of the system for all values of
the interaction strength. 
The Ansatz interpolates smoothly between 
the weak and the strong interaction case. It is worthwhile to
notice that the latter case cannot be decribed by the BP wave function
and, instead, it is described by the Thomas-Fermi approximation 
in Ref. \cite{BP}. Using this Ansatz,
correlation effects can be considered \cite{GIK}.

\section{Baym-Pethick trial wave  function and high temperature expansion}
\label{bps}
Baym-Pethick argued \cite{BP} that the initial condensate wave function
in the region of  weak and up to intermediate interactions can be well
approximated by a gaussian. Let us take a different point of view and
ask under which condition  this gaussian-like
profile of the initial condensate wave function
can be actually derived. 

If interactions between atoms are neglected, the physical system
is equivalent to the system of noninteracting harmonic oscillators.
In the latter case, one can show that the aspect ratio
in thermal equilibrium is a monotonically increasing function of
the inverse temperature $\beta=1/T$,
$$
R(\lambda,\beta)=\sqrt{\lambda \
\frac{(1/2)+ (e^{\beta\lambda\omega_\perp}-1)^{-1}}
{(1/2)+ (e^{\beta\omega_\perp}-1)^{-1}}} \cdot
$$
At high temperatures $R\approx 1$, and for low temperatures $R\approx
\sqrt{\lambda}$. In the latter case, the dominant contribution
to the aspect ratio is given by the ground state of the system
and reflects its anisotropy.

In the presence of the interactions, the Hamiltonian  of the system
can be written as 
\begin{equation}
H = \frac{\hbar^{2}}{2 m} \int dV \left[
\nabla \psi^{+} \nabla \psi +
a_{\perp}^{-4}(\rho^{2}+\lambda^2 z^2) \psi^{+}\psi +
4\pi l \psi^{+}\psi^{+}\psi\psi \right],
\label{H}
\end{equation}
where $\rho^2=x^2+y^2$, $a_\perp$ and $a_z$ are oscillator lengths,
and  $l$ is the s-wave scattering length  \cite{BP,Osc}.
Qualitatively, the temperature dependence of $R(\lambda,\beta)$
preserves the main features of the noninteracting case.
At high temperatures, the interaction is irrelevant and $R\approx 1$.
At low temperatures (for sufficiently small fraction of atoms
out of BEC), the HF approximation is justified and
the ground state wave function
(giving the main contribution to the aspect ratio) satisfies
the NSE. After rescaling of variables \cite{DS}, the NSE 
can be written as
\begin{equation}
\left[-\triangle + x^2+y^2+\lambda^2 z^2+ A |\psi ({\bf r})|^2 \right]
\psi({\bf r})= 2C \psi({\bf r}).
\label{nse}
\end{equation}
Here, $A= 8\pi l N/a_\perp$
characterizes the interaction strength, $N$ is the number
of particles in the condensate ($A\sim 520$ for $N\sim 5000$ \cite{Num}),
and $C=\mu/(\hbar\omega_\perp)>0$, $\mu$ being the chemical potential.
In the case of  large condensate fraction (strongly interacting case,
$A\gg 1$), the kinetic term can be neglected \cite{BP,DS} and the
ground state (normalized to unity) wave function is
given by the Thomas-Fermi approximation,
\begin{equation}
f^2({\bf r})=\frac{1}{A}\left(2C-x^2-y^2-\lambda^2 z^2\right) \Theta
\left(2C-x^2-y^2-\lambda^2 z^2\right),
\label{strong}
\end{equation}
where $2C=[15\lambda A/(8\pi)]^{2/5}$,
and $\Theta$ is the Heaviside step function.
The aspect ratio, $R(A,\lambda)$, is increasing function of $A$, and
the ground state solution (\ref{strong}) takes on its maximal possible value
($R=\lambda$) among  all ground state solutions to the NSE.
In the present case ($\lambda=\sqrt{8}$), this means that the
maximal effect of interactions is to raise the value of the aspect
ratio on $67\%$ with respect to the noninteracting case.
Moreover, as shown by Dalfovo and Stringari \cite{DS} the aspect ratio
for $A=520$ (corresponding to $N=5000$ atoms in BEC) is $R=2.3$,
i.e., $37\%$ higher than in the noninteracting case.

Let us now return to the BP wave function \cite{BP}.
In order to simplify the derivation of the BP
wave function from the high-temperature expansion,
let us  for a while, such as in Ref. \cite{BP}, neglect the
anisotropy of the oscillator potential \cite{If}.
Because the kinetic energy of particles
in the system is approximately 200 times smaller than
the characteristic interaction energy \cite{BP}, it is reasonable
to consider the expansion of the partition function of the
system, $Z(\beta, \mu)$, in powers of the kinetic term,
\begin{equation}
Z(\beta, \mu) =
\int D\psi^{+} (x,\tau )D\psi (x,\tau ) e^{S_0}
\sum_{n=0}^{\infty} \frac{[-\hbar^2/(2m)]^n}{n!}
\left(\int_{0}^{\beta} \int dV \ \nabla \psi^{+} \nabla \psi\right)^{n}
 \equiv \sum_{n=0}^{\infty} Z_{n}(\beta, \mu) \ .
\label{Z}
\end{equation}
Here $S_{0}$ is the ``unperturbed" action,
$$
S_{0} = \int_{0}^{\beta} \int dV
\left\{ \frac{\partial \psi^{+}}{\partial \tau}\psi
- \frac{\hbar^{2}}{2 m}
\left[\left(a_{\perp}^{-4}r^{2}  - \frac{2 m}{\hbar^{2}} \mu\right)
\psi^{+}\psi + 4\pi l \psi^{+}\psi^{+}\psi\psi \right] \right\} \ ,
$$
with fields satisfying the periodic boundary conditions,
$\psi^{+}(x,\beta)=\psi^{+}(x,0)$ and $\psi(x,\beta)=\psi(x,0)$.
Expansion (\ref{Z}) is the high temperature expansion and we are not
exactly in the BEC regime. However, the HF approximation is avoided.

In the first term (corresponding to $n=0$) of the sum in Eq. (\ref{Z})
 one finds a product of partition functions of the anharmonic
oscillators $\{H_{x}\}$, labelled by the space point $x$.
Using  the lattice approximation
$x=l (m_{1},m_{2},m_{3})\equiv l m$ with the scattering length
$l$ (the smallest length in the system) being the lattice spacing,
one has for field operators
$\psi^{+}(x) =\psi^{+}_{m}/\sqrt{l^3}$ and
$\psi(x) =\psi_{m}/\sqrt{l^3}$,
and
$$
Z_{0}(\beta, \mu) = \prod_{m} Z^{m}(\beta, \mu) .
$$
Here, $Z^{m}(\beta, \mu)$ is the partition function of the one-site
Hamiltonian $H_m$,
$$ H_{m} = \frac{\hbar^{2}}{2ml^2} \left[
\frac{l^4}{a_{\perp}^{4}}
(m_{1}^{2} + m_{2}^{2} + m_{3}^{2}) -
\frac{2 m l^2}{\hbar^{2}} \mu - 4\pi\right]
\psi^{+}(x)\psi(x) + 4\pi
\left(\psi^{+}(x)\psi(x)\right)^{2}.
$$
After redefining parameters,
$$
\mu \rightarrow \tilde{\mu}\equiv \frac{2ml^2}{\hbar^2}\mu, \quad
\beta\rightarrow \tilde{\beta}\equiv \frac{\hbar^2}{2ml^2}\beta, \quad
b_m = \frac{l^4}{a_\perp^4}
\,(m_1^2 + m_2^2 + m_{3}^{2})
$$
one has
$$
Z^{m}(\beta, \mu) = \sum_{k=0}^{\infty}
\exp\left(-4\pi\tilde{\beta} k^{2} +
\tilde{\beta}\tilde{\mu} k + 4\pi\tilde{\beta}  k -
\tilde{\beta}k b_{m}
\right) \ .
$$
Let us assume that the value of the chemical potential
$\tilde{\beta}\tilde{\mu}$ is negative and order of
10 (self-consistency
of this assumption will be shown below).  By substituting
$T \simeq  10^{-7}$K for the temperature, one has
$
 4 \pi \tilde{\beta} \simeq 12 \pi \times 10^{3},
$
and
\begin{equation}
Z^{m}(\beta, \mu) = 1 +
e^{\tilde{\beta}\tilde{\mu} - \tilde{\beta} b_{m}}
\left(1 + O(e^{-2.5\pi 10^{4}})\right) \simeq
1 + e^{\tilde{\beta}\tilde{\mu} - \tilde{\beta} b_{m}} \ .
\end{equation}
The resulting partition function,
$$
Z_{0}(\beta, \mu) = \prod_{m}
\left(1 + e^{\tilde{\beta}\tilde{\mu} - \tilde{\beta}
b_{m}}\right),
$$
leads to the distribution function $\langle
\psi^{+}(x)\psi(x)\rangle$ of
the Fermi-Dirac type,
$$
\langle\psi^{+}(x)\psi(x)\rangle =
\frac{1}{1 + e^{-\tilde{\beta}\tilde{\mu} +
\tilde{\beta} b_{m}}},
$$
with the chemical potential $\tilde{\mu}$
to be determined from
the normalization condition,
\begin{equation}
N = \int \frac{dV}{1 + e^{-\tilde{\beta}\tilde{\mu} +
\tilde{\beta} b_{m}}}\cdot
\label{N}
\end{equation}
Let $N \simeq 5000$ be the number of particles in the system.
Using that $l^4/a_{\perp}^{4} \simeq 2 \times
10^{-10}$, and, neglecting $1$ in the denominator of
Eq. (\ref{N}),
one gets
$
e^{\tilde{\beta}\tilde{\mu}} \simeq 1.3 \times 10^{-10} N/\pi^{3/2}
$
which implies
$
\tilde{\beta}\tilde{\mu} \simeq \ln \left( 10^{-7}\right)
\simeq -16.1,
$
in full accord with our assumption.

We cannot justify our treatment for low temperatures.
Nevertheless, under the assumption that our expansion
holds up to low temperatures,
the profile of the square root of the density function
$\psi_{0}(x)$ of the system turns out to be the BP trial wave
function \cite{BP},
$$
\psi_{0}(x) =
\sqrt{\langle\psi^{+}(x)\psi(x)\rangle}=\mbox{const} \times
\exp\left(-\frac{r^2}{2 \tilde{a}_{\perp}^{2}}\right),
$$
where
$
\tilde{a}_{\perp}  = a_{\perp}^2/\left(l\sqrt{\tilde{\beta}}\right)$.
The very fact that the BP wave function is the first order result
suggest that it may not describe BEC well. Therefore, the aspect ratio
obtained from the BP wave function may not be reliable.
Moreover, the profile of the BP function differs significantly
from  the exact ground state wave function calculated numericaly
in Refs.\cite{EDC,DS}.

\section{New analytical Ansatz}
\label{ans}
We shall show that the ground-state wave
function $f({\bf r})$ of the system
can be well described by a simple analytical Ansatz (cf. Fig. \ref{pl3}),
\begin{equation}
f^2 =\frac{1}{A}
W\left[A \exp(4C-\rho^2 - R^2(A,\lambda) z^2)\right],
\label{fe}
\end{equation}
where $W(x)$, defined as the principal branch (regular at the origin)
solution to the Eq. $We^W=x$ \cite{FSZ},
is  a standard MAPLE function \cite{MP}.
Constant $C$ in  Eq. (\ref{fe}) is to be determined from the normalization
condition, $\int f^2(r)\,d^3x=1$.
To approximate the ground-state solution to the NSE,
the value of $R$ should be supplied from the numerical
solution~\cite{DS}.
On the other hand, given experimental value of $R$,
our Ansatz can serve to reproduce the profile of the ground state.
\begin{figure}
\centerline{\epsfxsize=10cm\epsfbox{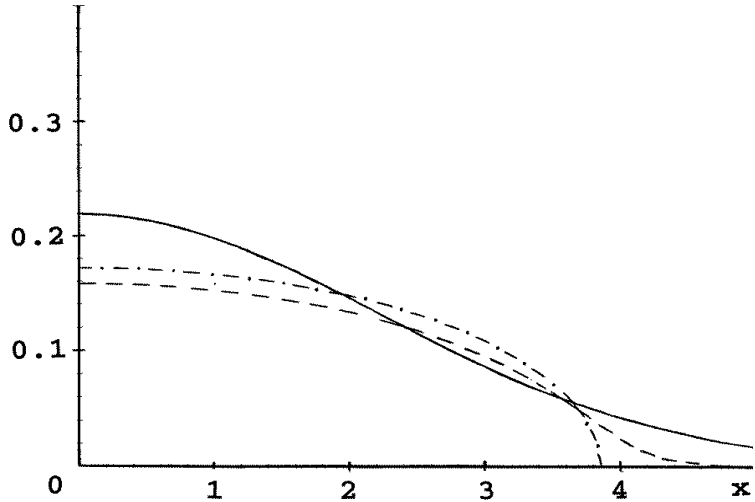}}
\caption{A comparison of the $x$-dependence  of the 
ground-state solution of Baym and Pethick
(solid line), strong limit (dot-dashed line) and our approximate
solution (dashed line).
The same values of the parameters were used as for 
Fig. \ref{pl3}.  
}
\label{pl1}
\end{figure}
\begin{figure}
\centerline{\epsfxsize=10cm \epsfbox{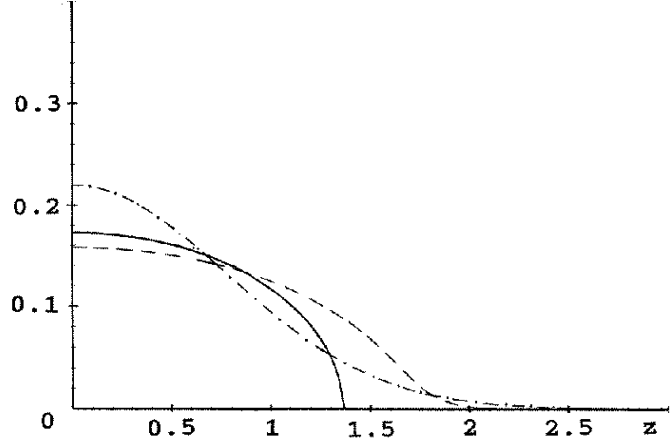}}
\caption{A comparison of the 
$z$-dependence of the ground-state solution of Baym and Pethick
(dot-dashed line), strong limit (solid line) and our approximate
solution (dashed line).}
\label{pl2}
\end{figure}

Eq. (\ref{nse}) is a nonlinear equation and, in the present case, no exact
solutions are known except for the two limiting cases, $A=0$ and
$\triangle f\ll Af^3$.
In what follows, we shall construct our Ansatz to reproduce correctly 
the two limiting cases and to interpolate smoothly between them 
as interaction changes. 
Note that in the strong interaction limit
the Baym-Pethick trial wave-function cannot be used at all
and the Thomas-Fermi approximation was used in Ref. \cite{BP}
in this limit.
The main point in our derivation is
to use instead of the $2$nd order differential equation
(\ref{nse}) a set of {\em first} order differential equations which
reproduce correctly both the noninteracting limit ($A=0$)
and the strongly interacting limit ($\triangle f\ll Af^3$).
In this sense, our Ansatz will be exact in the first derivatives.
In the noninteracting limit, Eq.\ (\ref{nse}) reduces to
the stationary Schr\"{o}dinger equation for an anisotropic
oscillator and the (normalized to unity) ground state  wave function is
\begin{equation}
f({\bf r})=\frac{\lambda^{1/4}}{\pi^{-3/4}}\,\exp\left[
-\frac{1}{2}\left(
x^2+y^2+\lambda z^2\right)\right],
\label{non}
\end{equation}
with $C=2+\lambda$ and, in agreement with our previous discussion,
$R=\sqrt{\lambda}$.
One notice that
$f({\bf r})$ satisfies the set of first order differential equations,
\begin{equation}
\partial_1 f=-xf,\hspace*{1cm} \partial_2 f=-yf,\hspace*{1cm}
\lambda^{-1}\partial_3 f= R^{-2}(\lambda)\partial_3 f=- zf.
\label{1cas}
\end{equation}
On the other hand, if $\triangle f\ll Af^3$, $Af^3/\lambda$,
the (normalized to unity) ground-state wave function is given 
by the Thomas-Fermi approximation [see Eq. (\ref{strong})], and
\begin{equation}
Af^2\partial_1 f\approx -xf,\hspace*{1cm} Af^2\partial_2 f\approx
-yf,\hspace*{1cm}
\lambda^{-2} Af^2\partial_3 f =
R^{-2}(\lambda) Af^2\partial_3 f\approx -  z f.
\label{2cas}
\end{equation}
Now, first order differential equations (\ref{2cas}) can be combined into
the  variational principle,
$E[f]=\int \sum_j P_jP_j\,d^3{\bf r}$,
where
$P_1 =\left[(1+Af^2)\partial_1 +x\right]f$,
$P_2 =\left[(1+Af^2)\partial_2 +y\right]f$,
$P_3 =\left[R^{-2}(A,\lambda)(1+Af^2)\partial_3 + z\right]f$.
The actual form of $P_3$ in the asymmetric case is fixed by
the requirement to make simultaneous integration of the first order
differential equations (\ref{first})  possible.
The variational principle implies the following set of first order
equations for the ground state wave function,
\begin{equation}
\partial_\rho f= -\frac{\rho f}{1+Af^2},\hspace*{1cm}
\partial_3 f= - R^2(A,\lambda)\frac{ z f}{1+Af^2}.
\label{first}
\end{equation}
By integrating Eqs. (\ref{first}) one obtains that
$$
f e^{Af^2/2}=\exp\left[2C-\rho^2/2-R^2(A,\lambda)z^2/2\right],
$$
from which our Ansatz (\ref{fe}) follows immediately.
Provided that $\rho^2+R^2(A,\lambda) z^2\leq 4C$,
$f$ can be found explicitly using successive iterations,
\begin{equation}
f^2= \frac{1}{A}\ln\frac{s^2}{f^2} =  \frac{1}{A}\ln
\frac{s^2}{
\frac{1}{A}\ln \frac{s^2}{\frac{1}{A}\ln \frac{s^2}{\ldots}}} >0,
\label{sol2}
\end{equation}
where $s=\exp\left[2C-\rho^2/2- R^2(A,\lambda) z^2\right]$.
Obviously,  $f({\bf r})$ given by Eq.\ (\ref{fe}) reproduces correctly
the ground state wave function both in the noninteracting
limit ($A=0$) and in the strongly interacting limit
($Af^2\gg 1$), and  interpolates smoothly between the two limiting cases
in the intermediate region (see Figs. \ref{pl1} and\ref{pl2}).
 One can verify that
if $Af^2\gg 1$  then $\triangle\,f\sim -[1/(Af^2)^2]
(x^2+y^2+ R^2(A,\lambda) z^2 +3Af^2)f$,
and the kinetic term
is suppressed by the factor $(Af^2)^{-2}$ with respect to the
remaining terms in (\ref{nse}).

Our Ansatz can substitute for the BP trial wave function and can play
the role of a new trial wave function in various variational
calculations.
Given experimental values of the aspect ratio, our Ansatz can be effectively
applied to describe the initial BEC wave function and
to calculate all relevant properties of the initial BEC.
We believe that the derivation of our Ansatz can be extended to
deal with excited states too.

\section{Discussion and conclusions}
%%%%%%%%
In the paper, only  the properties of the initial condensate 
were considered.  To find the  connection with  experiment,
it is necessary to discuss properties of the 
system during the transition from the strong trap
to the weak trap and its subsequent ballistic expansion from the weak trap.
Obviously, characteristics of the system will change after 
the expansion  and will strongly depend on 
the condition of the expansion (i.e., whether it is adiabatic or abrupt). 
Nevertheless, the present study allows us 
to give an upper bound for the aspect ratio of the condensate after its
expansion directly from the strong trap, i. e., 
in the absence of the intermediate weak trap,
as it took place in \cite{Exp1}: {\em the  aspect ratio of the final 
system is always lower than that  calculated for  the initial condensate}. 
Indeed, after the expansion, 
(i) there is no more anisotropic potential
applied, (ii) the self-interaction of bosons, which gives rise
to the increase of the aspect ratio with respect to the 
noninteracting case, is decreasing. 
Note that if one can measure  the aspect ratio 
directly after the expansion from the strong trap
as it should be done in future experiments,
it would be possible to estimate the real number of the particles in the
condensate.

We want to emphasize that, 
as it can be found from comparison of the results 
obtained from the BP wave function, the Thomas-Fermi approximation, 
and from the exact numerical solution, 
the aspect ratio may be a very sensitive characteristic
of the wave function and may change considerably even when other
characteristics are not changed by a perturbation (such as an external
potential or an interaction). That is why we expect that taking into account correlation effects may lead to the considerable changes in the aspect ratio although for other characteristics the Hartree-Fock approximation
will give correct results. We will consider this question in details
in a forthcoming paper \cite{GIK}.

Summarizing, given the value of the aspect ratio, both the
profile of the ground state and the quality of  BEC can be estimated.
This allows one to estimate the number of  particles 
in the initial condensate. 
We showed that the Baym-Pethick trial wave-function 
(i) is only the first order approximation in the high temperature 
expansion for the system and (ii) does not describe the condensate 
wave function accurately even for weak and intermediate interactions.
Note that in the strong interaction limit
the Baym-Pethick trial wave-function cannot be used at all
and instead the Thomas-Fermi
approximation was used in Ref. \cite{BP}.
In order to describe the ground state of the initial condensate in 
the whole range of the interparticle interactions,
we proposed a simple analytical Ansatz which, in contrast to the 
BP trial wave function, reproduces correctly both the weak and the
strong interaction limit and interpolates smoothly between
the two limiting cases as interaction changes.

We want to thank  J.M.F. Gunn for raising our interest in the problem
and for many encouraging discussions. We are also grateful to 
K. Burnett, F. Dalfovo, M. Holland, and C. Pethick for discussion, 
and P. Cooper for careful reading of the manuscript.
This work was partially supported (K.I.) by the Grant of Russian Fund of
Fundamental Investigations N 95-01-00548,  the UK EPSRC Grant GR/J35221, and
(A.M.) by  the UK EPSRC Grant GR/J35214, and by the Grant Agency of 
the Czech Republic under Project No. 202/93/0689.


\begin{thebibliography}{99}
\bibitem{Osc}The $s$-wave scattering length can be estimated as
$l \sim 100 a_{0}$ \cite{a}, where
$a_{0}$ is the Bohr radius (i.e., $l \sim 5 \cdot 10^{-7} \mbox{cm}$).
\bibitem{Num}Here the value $N\sim 5000$ is chosen in order
to  make a comparison our Ansatz with the numerical solution of
Ref.~\cite{DS}) possible.
\bibitem{If}If $\lambda\neq 1$, one can rescale coordinate $z$
in Eq. (\ref{H}) according to
$z\mapsto \bar{z}= \sqrt{\lambda} z$ which implies that
$-\partial_z^2 +\lambda^2 z^2\mapsto
\lambda\left(-\partial_{\bar{z}}^2 +\bar{z}^2\right).$
Therefore, the only effect of the asymmetry is that the $s$-wave
scattering length $l$
becomes effectively $l/\lambda$ in the $z$-direction.
\end{thebibliography}
\end{document}